\documentclass[12pt]{iopart}
\usepackage{epsf}
\usepackage{graphicx}
\def\gtsim {\lower .1ex\hbox{\rlap{\raise .6ex\hbox{\hskip .3ex
        {\ifmmode{\scriptscriptstyle >}\else
                {$\scriptscriptstyle >$}\fi}}}
        \kern -.4ex{\ifmmode{\scriptscriptstyle \sim}\else
                {$\scriptscriptstyle\sim$}\fi}}}

\def\ltsim {\lower .1ex\hbox{\rlap{\raise .8ex\hbox{\hskip .3ex
        {\ifmmode{\scriptscriptstyle <}\else
                {$\scriptscriptstyle <$}\fi}}}
        \kern -.4ex{\ifmmode{\scriptscriptstyle \sim}\else
                {$\scriptscriptstyle\sim$}\fi}}}
                
\begin{document} 

\title[Neutrino Coherent Scattering Rates at Direct Dark Matter Detectors]
{Neutrino Coherent Scattering Rates at Direct Dark Matter Detectors}

\author{Louis E. Strigari} 

\address{Kavli Institute for Particle Astrophysics and Cosmology, Stanford University, Stanford, CA 94305, USA}
\ead{strigari@stanford.edu}

\begin{abstract} 
Neutrino-induced recoil events may constitute a background to 
direct dark matter searches, particularly for those detectors that strive to
reach the ton-scale and beyond. 
This paper discusses the expected neutrino-induced background spectrum 
due to several of the most important sources, including solar, atmospheric, and diffuse supernova neutrinos.
The largest rate arises from $^8$B produced solar neutrinos,
providing upwards of $\sim 10^3$ events per ton-year over all recoil energies
for the heaviest nuclear targets. However the majority of these $^8$B events are
expected to be below the recoil threshold of modern detectors. The remaining
neutrino sources are found to constitute a background to the WIMP-induced recoil 
rate only if the WIMP-nucleon cross section is less than $10^{-12}$ pb. Finally 
the sensitivity to diffuse supernova neutrino flux for non-electron neutrino flavors
is discussed, and projected flux limits are compared with existing flux limits. 
\end{abstract} 

\maketitle 

\section{Introduction} 
Weakly-Interacting Massive Particles (WIMPs) are a leading 
candidate for the dark matter that constitutes $\sim 23\%$ 
of the mass density of the Universe~\cite{Jungman:1995df}. 
A promising means to search for WIMPs is through direct interactions
in underground detectors, whereby the struck nucleus recoils
coherently and its kinetic energy is deposited into 
scintillation light. At the present, detectors
constrain the spin-independent WIMP-nucleon cross
section to be $\ltsim \, 9 \times 10^{-44}$ cm$^2$
for WIMPs in the mass range $\ltsim \, 100$ GeV
~\cite{Angle:2008we,Ahmed:2008eu}.  
Future detectors are expected to reach the ton-scale and beyond, 
providing the exciting prospect of covering much of the 
model parameter space for leading dark matter candidates~\cite{Roszkowski:2007fd}. 

Present direct detection experiments report upper limits
on the WIMP mass and cross section under the
assumption of unknown backgrounds~\cite{Yellin:2002xd}. 
As the sizes of these detectors continue to increase, 
backgrounds that are beyond the sensitivity limit of modern detectors 
will reveal themselves, necessitating a detailed understanding of 
these background spectra and new techniques for 
their reduction. 
Due to the nature of the detected events, background subtraction 
requires
understanding the rate and spectral shape of any signal that leads to 
a coherent nuclear recoil event in the energy 
window where a dark matter signal appears. 

In his seminal work over thirty years ago, Freedman~\cite{Freedman:1973yd} pointed
out that the neutrino-nucleon neutral current interaction
leads to a coherence effect, whereby the neutrino-nucleus elastic scattering cross section
is enhanced and scales approximately as the square
of the number of neutrons in the nucleus. Coherent recoils are
expected if the three-momentum imparted to the nucleus, $q$, obeys the
inequality $qR \ltsim \, 1$, where $R$ is the radius of the nucleus. 
For typical nuclear radii, 
coherent scattering leads to nuclear recoils in the range of a few keV 
for incoming neutrinos with energies in the range $\sim 1-100$ MeV.  
Though neutrino-nucleus coherent scattering remains a fundamental 
prediction of the Standard Model, this process has yet to be 
observed experimentally~\cite{Scholberg:2005qs}. 

Solar, atmospheric, and reactor neutrinos 
may induce observable nuclear recoil events
~\cite{Cabrera:1984rr}. 
Monroe and Fisher~\cite{Monroe:2007xp} have recently 
determined the coherent scattering rates in future dark matter
detectors due to solar, atmospheric, and geo-neutrinos, in
particular identifying $^8$B solar neutrinos as the source
with the largest event rate. For detectors
at the $\sim$ ton-yr exposure scale, $^8$B neutrino-induced
recoils are significant for recoil energies
$\ltsim \, 3$ keV for heavy targets such as Xe and $\ltsim \, 30$ keV
for lighter targets such as Flourine and Carbon. 
Assuming a $\sim 100$ GeV WIMP, $^8$B neutrino-induced
recoils constitute a background to the WIMP recoil spectrum if the WIMP-nucleon 
cross section is $\ltsim \, 10^{-10}$ pb~\cite{Monroe:2007xp,Vergados:2008jp}. 

The goal of this paper is to provide a survey of the expected event rate recoil spectra
from several important astrophysical neutrino sources. Solar, atmospheric, 
and diffuse supernova neutrinos are studied; geo-neutrinos are not considered
since their signal is dwarfed by the solar neutrino signal over a similar energy 
range~\cite{Monroe:2007xp}. In addition to providing an
updated calculation of the solar and atmospheric 
neutrino fluxes for a variety
of often-discussed nuclear targets, the
first predictions for the recoil energy spectrum 
of the {\it Diffuse Supernova Neutrino Background 
(DSNB)} are provided. 
If the $^8$B neutrinos can be fit, or if the dominant component
to their recoil spectrum lies below the analysis energy threshold, 
astrophysical neutrino 
backgrounds will only contaminate a WIMP-nucleon recoil signal if the WIMP-nucleon 
cross section is $\ltsim \, 10^{-12}$ pb. 
It is additionally shown that, 
with current exposures, detectors should have sensitivity to 
reduce the current upper limit on 
the $\nu_\mu$ and $\nu_\tau$ component of {\it DSNB} flux by more
than an order of magnitude.

\section{Neutrino-Induced Recoil Spectra} 

\subsection{Neutrino-Nucleus Coherent Scattering} 
The neutrino-nucleus scattering cross section per recoil kinetic energy $T$ is
~\cite{Freedman:1977xn} 
\begin{equation}
\frac{d\sigma(E_\nu,T)}{dT} = \frac{G_f^2}{4 \pi}Q_w^2 M
\left(1-\frac{MT}{2E_\nu^2}\right)F(Q^2)^2. 
\label{eq:coherent}
\end{equation} 
Here $Q_w = N - (1-4 \sin^2 \theta_w)Z$ is the weak nuclear charge, 
where $N$ is the number of neutrons and $Z$ is the number of protons. 
The mass of the nucleus is $M=A M_N$, where $A = N+Z$ is the mass
number and $M_N = 931$ MeV. From kinematics the recoil energy 
is related to the neutrino three-momentum via $q^2 = 2MT$, 
and the maximum three-momentum transfer is a function of the neutrino
energy, $q_{max}^2 = 4E_\nu^2$. The range of recoil energy for
the struck nucleus varies between 0 and $T_{\rm max}$, where 
\begin{equation} 
T_{\rm max} = \frac{2E_\nu^2}{M+2E_\nu}. 
\label{eq:Tmax}
\end{equation} 
Taking the limit $M \gg 2E_\nu$, which will always be valid for
the results presented here, the minimum neutrino energy for a fixed recoil 
kinetic energy $T$ is $E_\nu = \sqrt{M T/2}$. 

The form factor, $F(Q^2)^2$, is a function of the 4-momentum, 
$Q$, and measures the distribution of nucleons 
within the nucleus and thus the deviations from coherence at high
recoil energies, defined by 
momentum transfers $qR\,\gtsim\,1$. The form factor at zero momentum 
transfer scales as $F(0)^2 \rightarrow 1$ for $qR \ll 1$. To approximate
the effects of the form factor, 
the analysis in this paper uses the Helm form factor as parametrized in
Lewin and Smith~\cite{Lewin:1995rx}. 
As in the case of WIMP-nucleon recoils, the form factor has the 
primary effect of reducing the cross section for large momentum 
transfers.  

\subsection{Neutrino Fluxes} 
Since the coherent scattering process is ``flavor blind,"  
mixing angles which connect neutrino weak flavor eigenstates to 
mass eigenstates need not be considered in flux calculations.
For all of the sources, 
the energy spectrum for the weak flavor eigenstates
at the production sites are used. 

{\it Solar Neutrinos} - The calculations
in this paper consider two sources of solar neutrinos, 
so-called  $^8$B and {\it hep} neutrinos. These neutrinos come
from the reactions $^8$B $\rightarrow ^7$Be$^* + e^+ + \nu_e$
and $^3$He $+ p \rightarrow ^4$He $+ e^+ + \nu_e$, which 
occur in $2 \times 10^{-2}\%$ and $2 \times 10^{-5}\%$ of the terminations
of the solar {\it pp} chain, respectively
~\cite{Bahcall1989}. The remaining sources of
solar neutrinos induce recoils with energies and
rates below that of the $^8$B induced spectrum. For example, 
neutrinos produced directly in the 
{\it pp} reaction only induce 
recoils up to a maximum kinetic energy of 
$\sim 0.03$ keV~\cite{Monroe:2007xp}. 

{\it Atmospheric Neutrinos}--
Cosmic ray collisions in the atmosphere produce a primary flux
of $\nu_e$, $\bar{\nu}_e$, $\nu_\mu$, and $\bar{\nu}_\mu$
~\cite{Gaisser:2002jj}, with a spectrum that 
extends up to energies of $\sim 100$ GeV. At these high
energies, the loss of coherence in the scatterings must be
accounted for.  
On the one hand, the coherent scattering cross section
scales as the neutrino energy squared, favoring high 
neutrino energies. On the other hand, the form factor
suppresses scatterings for high momentum transfers, 
which largely result from the highest energy neutrinos 
in the flux distribution. Convolving both of these factors
means that detectors are mainly sensitive to
the atmospheric neutrino flux $\ltsim \, 100$ MeV. 
A more exact upper cut-off depends on the detector
target; the larger the mass number of the nucleus,
the more strongly the event rate
is suppressed by the form factor  
at the highest detectable recoil energies. For the 
atmospheric neutrino flux, 
the calculations in this paper use the published 
Fluka results for the Kamioka location, which have been 
determined down to neutrino energies 
$\sim 10$ MeV~\cite{Battistoni:2005pd}.
At energies of interest for this calculation, $\ltsim \, 100$ MeV, 
there is a $\sim 20\%$ uncertainty in the normalization of the flux,  
primarily due to geo-magnetic effects. 

{\it Diffuse Supernova Neutrinos}-
Core-collapse supernova produce a burst of $\sim 10^{58}$ neutrinos 
of all flavors per explosion; the past history of all of these explosions
produce a {\it Diffuse Supenova Neutrino Background (DSNB)}.  
The {\it DSNB} flux is a convolution of the core-collapse supernova rate
as a function of redshift with the neutrino spectrum per supernova. 
The core-collapse rate is derived from the star-formation 
rate and stellar initial mass function; the calculations in the paper 
follow the recent results of Horiuchi et al.~\cite{Horiuchi:2008jz}. The neutrino spectrum 
of a core-collapse supernova is believed to be similar to 
a Fermi-Dirac spectrum, with temperatures in the range 3-8 MeV
~\cite{Keil:2002in}. The calculations in this paper assume the following temperatures
for each neutrino flavor:  $T_{\nu_e} = 3$ MeV, $T_{\bar{\nu}_e} = 5$ MeV, 
and  $T_{\nu_x} =$ 8 MeV. Here $T_{\nu_x}$ represent the four flavors
$\nu_\mu$, $\bar{\nu}_\mu$, $\nu_\tau$, and $\bar{\nu}_\tau$. 
Due to the $\sim E_\nu^2$ scaling of the total cross-section in Eq.~\ref{eq:coherent}
(integrated over all recoil energies), 
the flavors with the largest temperature dominate the event rate. 

There are strong upper limits on the $\bar{\nu}_e$ component of
the {\it DSNB} flux from Super-Kamiokande, 
$< 1.2$ cm$^{-2}$ s$^{-1}$
~\cite{Malek:2002ns}. This flux limit constrains the
high energy tail of the {\it DSNB} flux, specifically neutrino energies $\gtsim \,18$ MeV. 
More recently, the SNO collaboration
has placed an upper limit on the $\nu_e$ component of the flux of
$\sim 70$ cm$^{-2}$ s$^{-1}$ for neutrinos in the energy range of 
$22.9-36.9$ MeV~\cite{Aharmim:2006wq}.  Though weaker than
the $\bar{\nu}_e$ limit, the SNO $\nu_e$ limit does probe the
parameter space for exotic core-collapse models
~\cite{Beacom:2005it}. An indirect bound on the $\nu_e$ flux
that rivals the present limit on the $\bar{\nu}_e$ flux follows from
oscillation conditions~\cite{Lunardini:2006sn}, and stronger direct bounds
on the $\nu_e$ flux may be realized in future Argon TPC detectors
~\cite{Cocco:2004ac}. 
The strongest bound on $\nu_x$ flavors comes from the 
Mt. Blanc experiment~\cite{Aglietta:1992yk}, which quotes an
upper limit on the $\nu_x$ flux of $\sim 10^7$ cm$^{-2}$ s$^{-1}$ for 
neutrino energies $\gtsim \, 20$ MeV. This flux limit was set based on the 
neutral current interaction with $^{12}$C. It has additionally been proposed 
that Super-Kamiokande can set an improved limit on the  
$\nu_x$ flux~\cite{Lunardini:2008xd}. 
Below the prospects for improving on these $\nu_x$ flux 
limits are discussed.

In addition to the {\it DSNB}, detectors sensitive to low energy nuclear recoils may be used to
detect a supernova burst in the Galaxy~\cite{Horowitz:2003cz}.
For exposures on the scale of $\sim$ ton-yr, there are expected to 
be of order 10 events, summed over all flavors for a supernova at 
10 kpc~\cite{Akerib:2006ks}. The detection of a supernova burst via 
a recoiling nucleus would provide a unique window into supernova
physics, complementing the detection of proton recoils
~\cite{Beacom:2002hs} as the best means to study the $\nu_x$ component of the neutrino flux. 

\begin{figure}[htbp]
\begin{center}
\includegraphics[height=7cm]{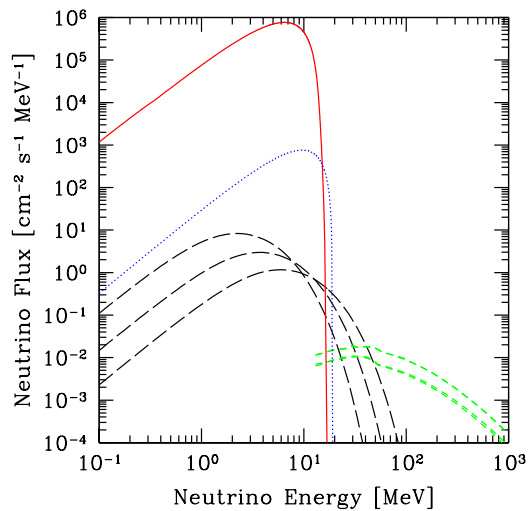}
\caption{Neutrino fluxes for astrophysical sources that constitute the dominant backgrounds 
to WIMP recoil signals. From top to bottom on the left, the fluxes are solar $^8$B (solid, red) 
and {\it hep} (dotted, blue), {\it DSNB} (long-dash, black) with temperatures of 3 
5, and 8 MeV (corresponding to 
$\nu_e$, $\bar{\nu}_e$, and $\nu_x$, respectively, where $\nu_x$ symbolizes
muon and tau neutrinos and the respective antiparticles). The short-dashed (green)
curves at the highest energies are the atmospheric neutrino fluxes, plotted down to 
the lowest energy in the calculation of Ref.~\cite{Battistoni:2005pd}. From 
bottom to top on the right, the fluxes are for $\nu_e$, $\bar{\nu}_e$, then 
$\nu_\mu$, $\bar{\nu}_\mu$. 
\label{fig:fluxes}
}
\end{center}
\end{figure}

Figure~\ref{fig:fluxes} shows the neutrino flux spectra for solar, 
atmospheric, and {\it DSNB}
sources. In terms of integrated flux, the $^8$B flux is by far dominant,
nearly three orders of magnitude larger 
than the next closest {\it hep} rate. However, as the $^8$B flux is confined to 
energies $\ltsim \, 16$ MeV, these will produce lower energy recoils 
in comparison to the higher energy {\it DSNB} and atmospheric components. A 
convolution between the flux and cross section is required to obtain
the recoil rate as a function of energy; the goal of the
following section is 
to determine these respective rates. 

\begin{figure*}[htbp] 
\begin{center}
\begin{tabular}{cc}
\includegraphics[height=7cm]{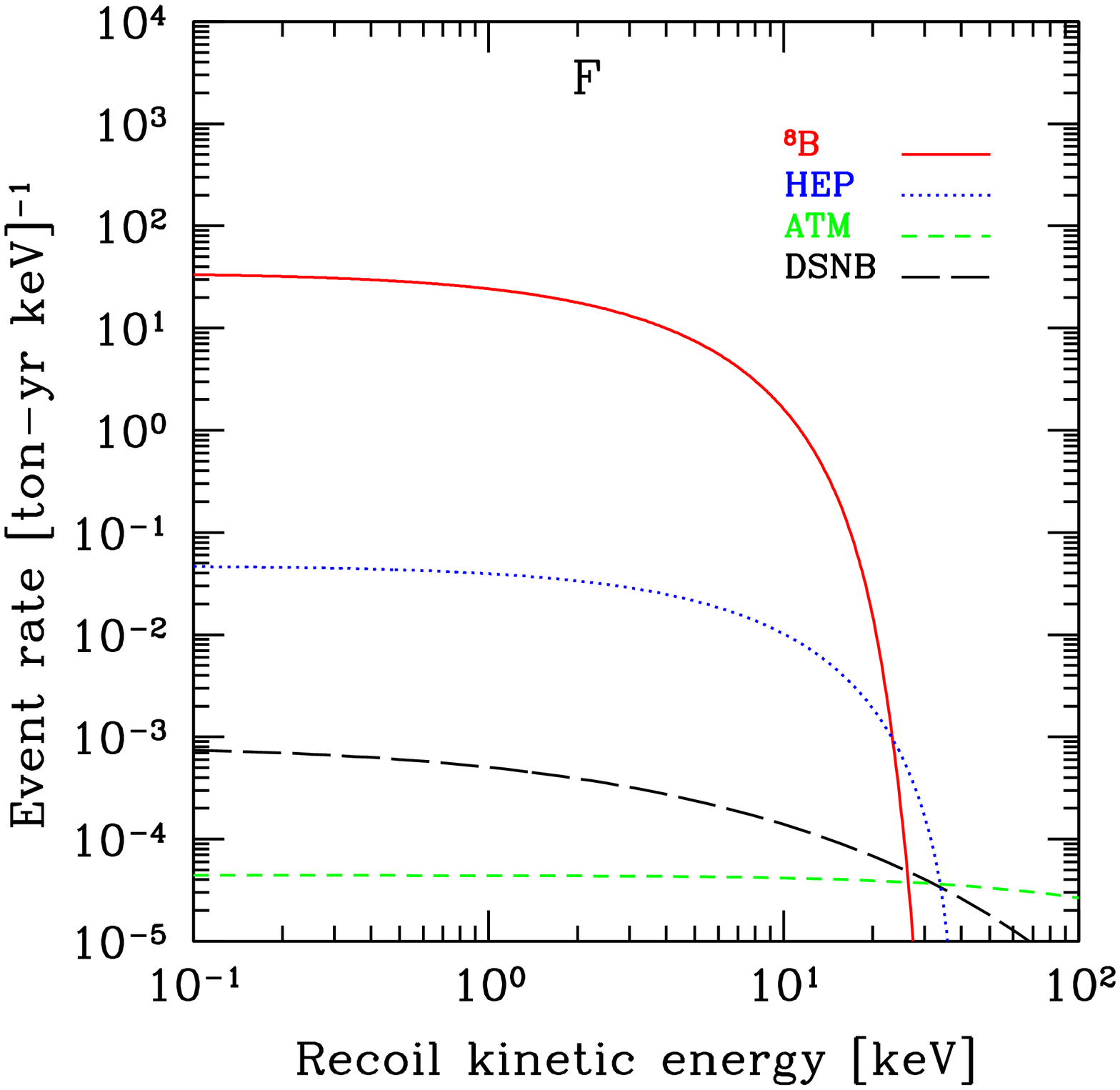} &
\includegraphics[height=7cm]{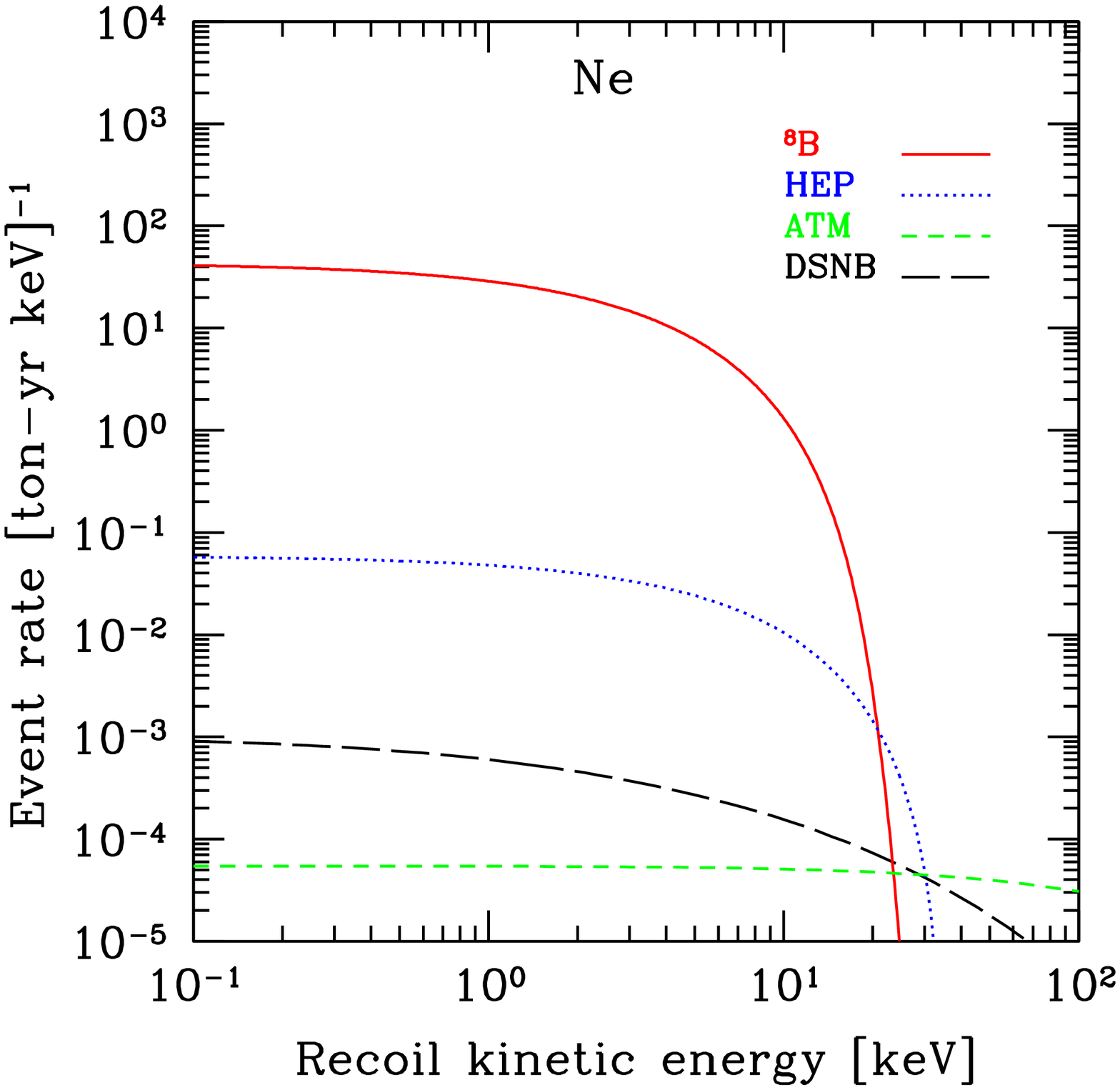} \\
\includegraphics[height=7cm]{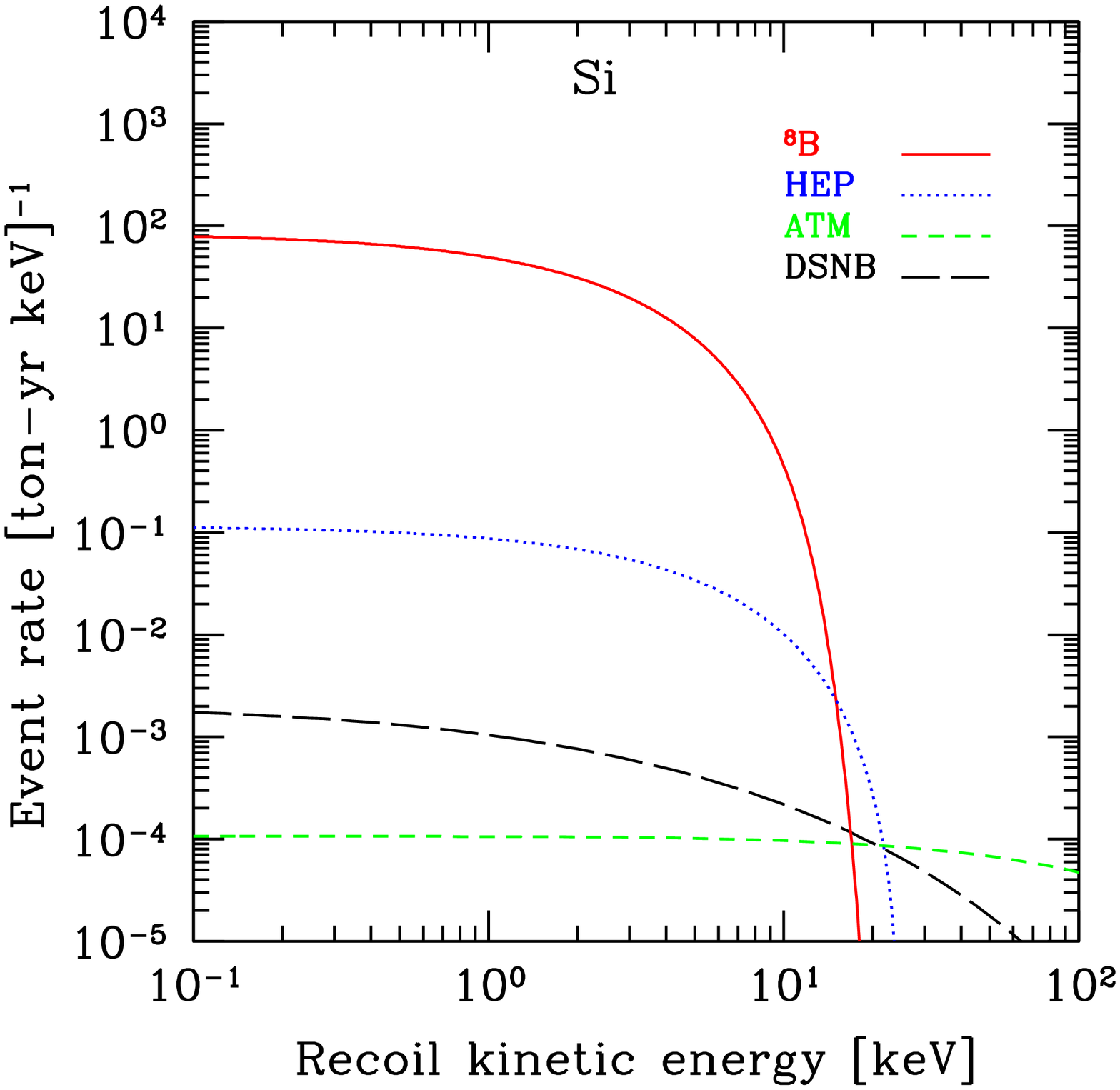} &
\includegraphics[height=7cm]{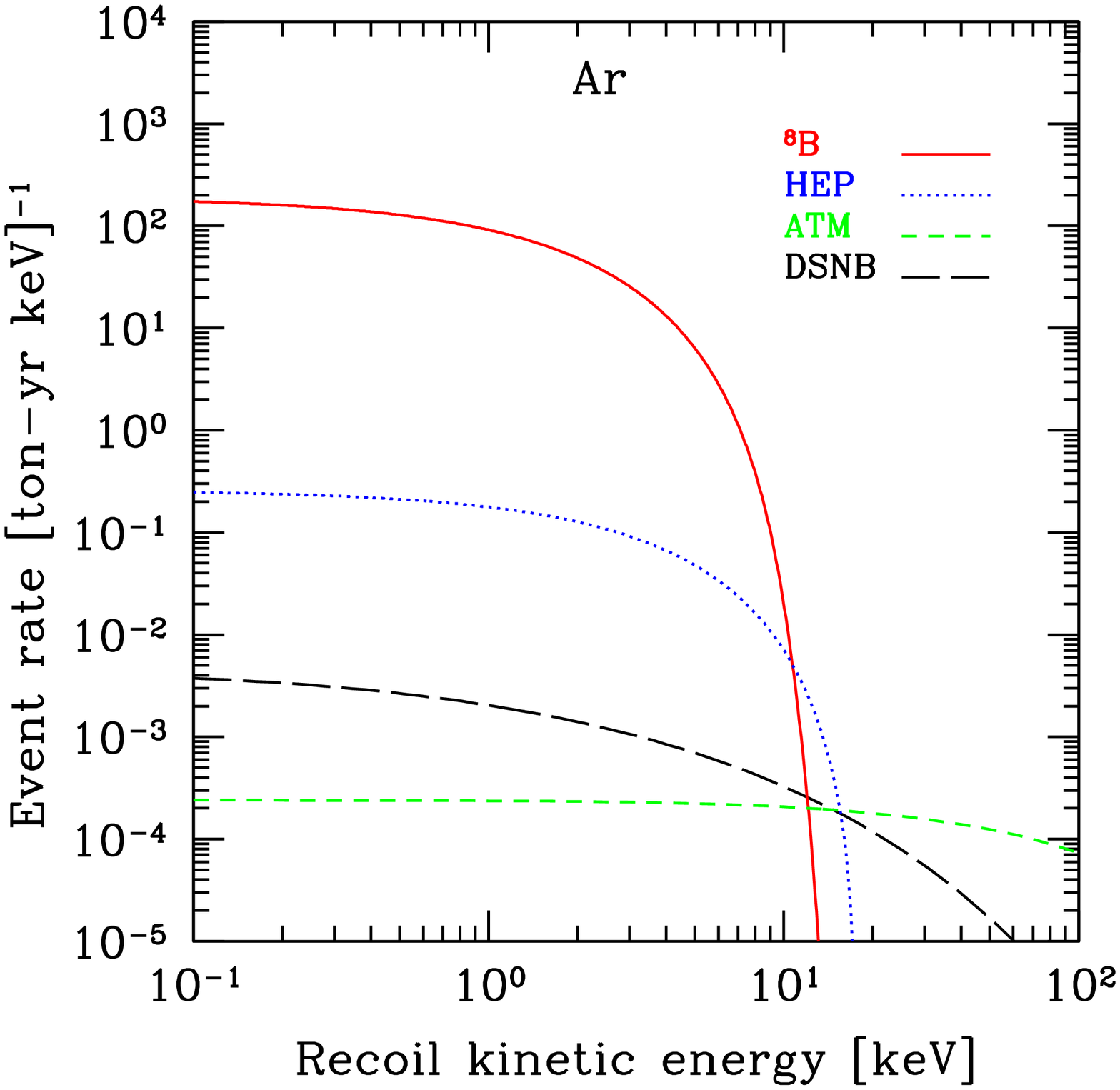} \\
\includegraphics[height=7cm]{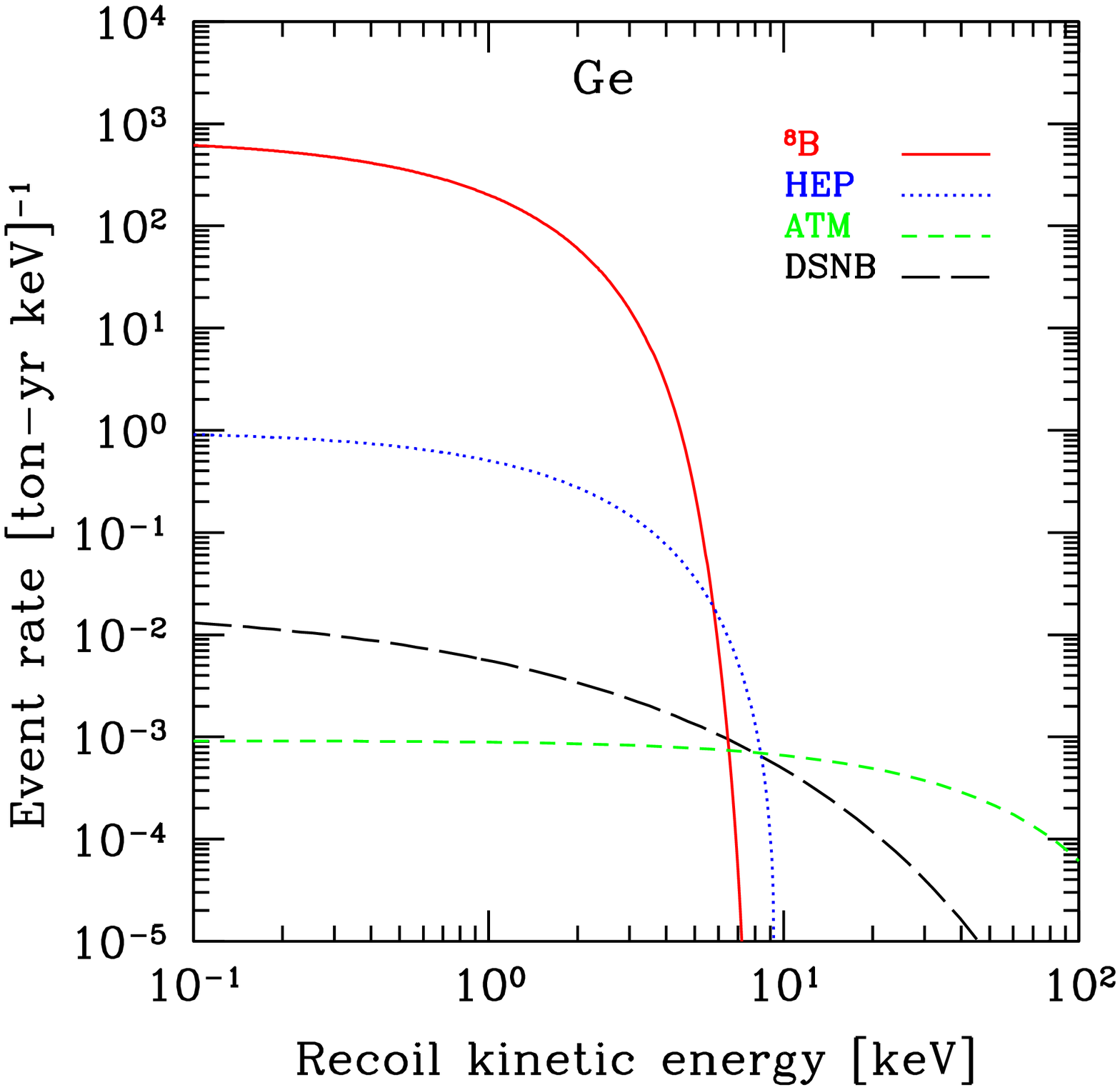} & 
\includegraphics[height=7cm]{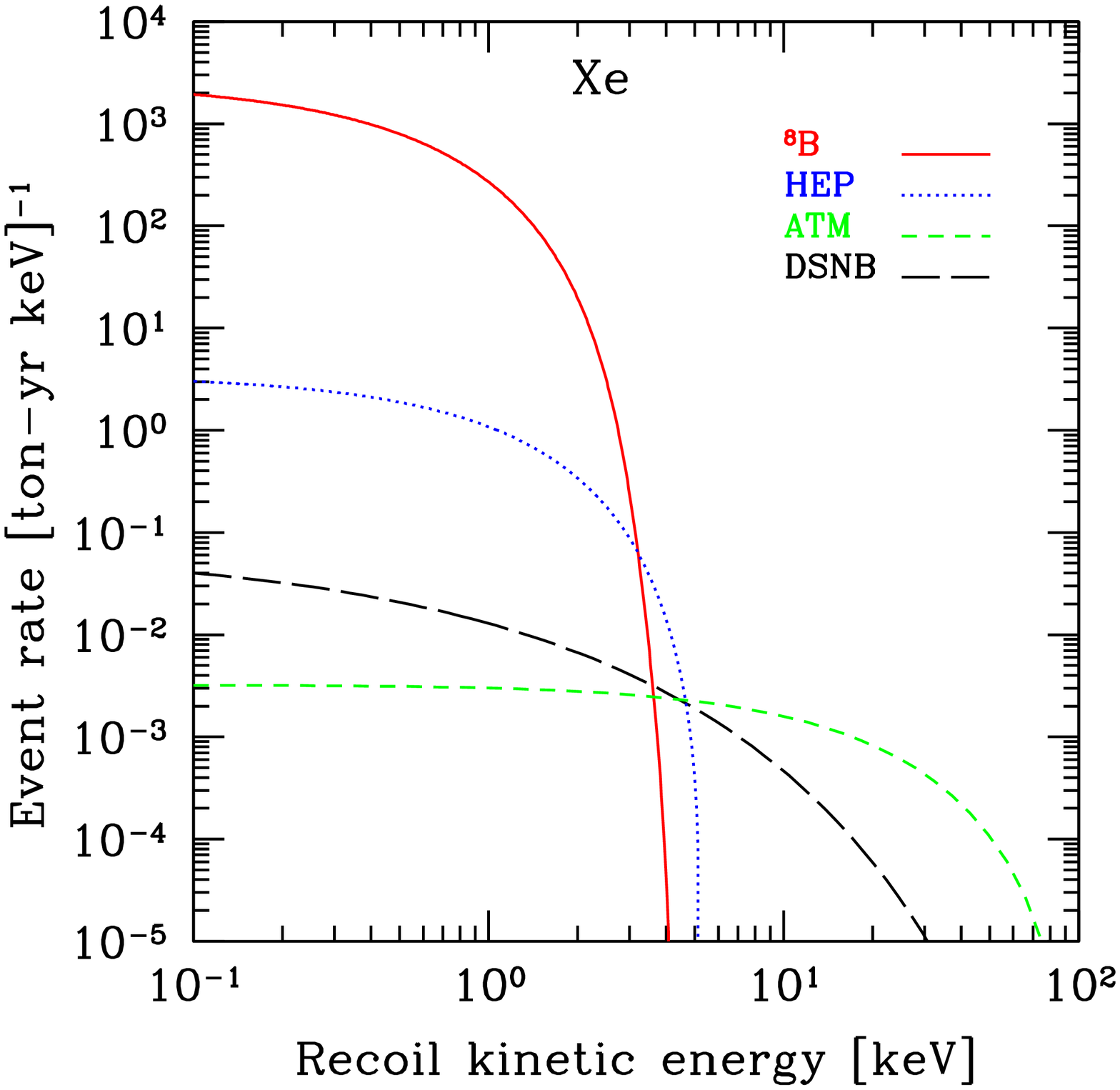} \\ 
\end{tabular}
\caption{Event rate per recoil kinetic energy 
for six target nuclei. For both the diffuse supernova and atmospheric event rates, the sum
of all contributing neutrino flavors are shown. 
\label{fig:rates}
}
\end{center}
\end{figure*}
\begin{figure*}[htbp] 
\begin{center}
\begin{tabular}{cc}
\includegraphics[height=7cm]{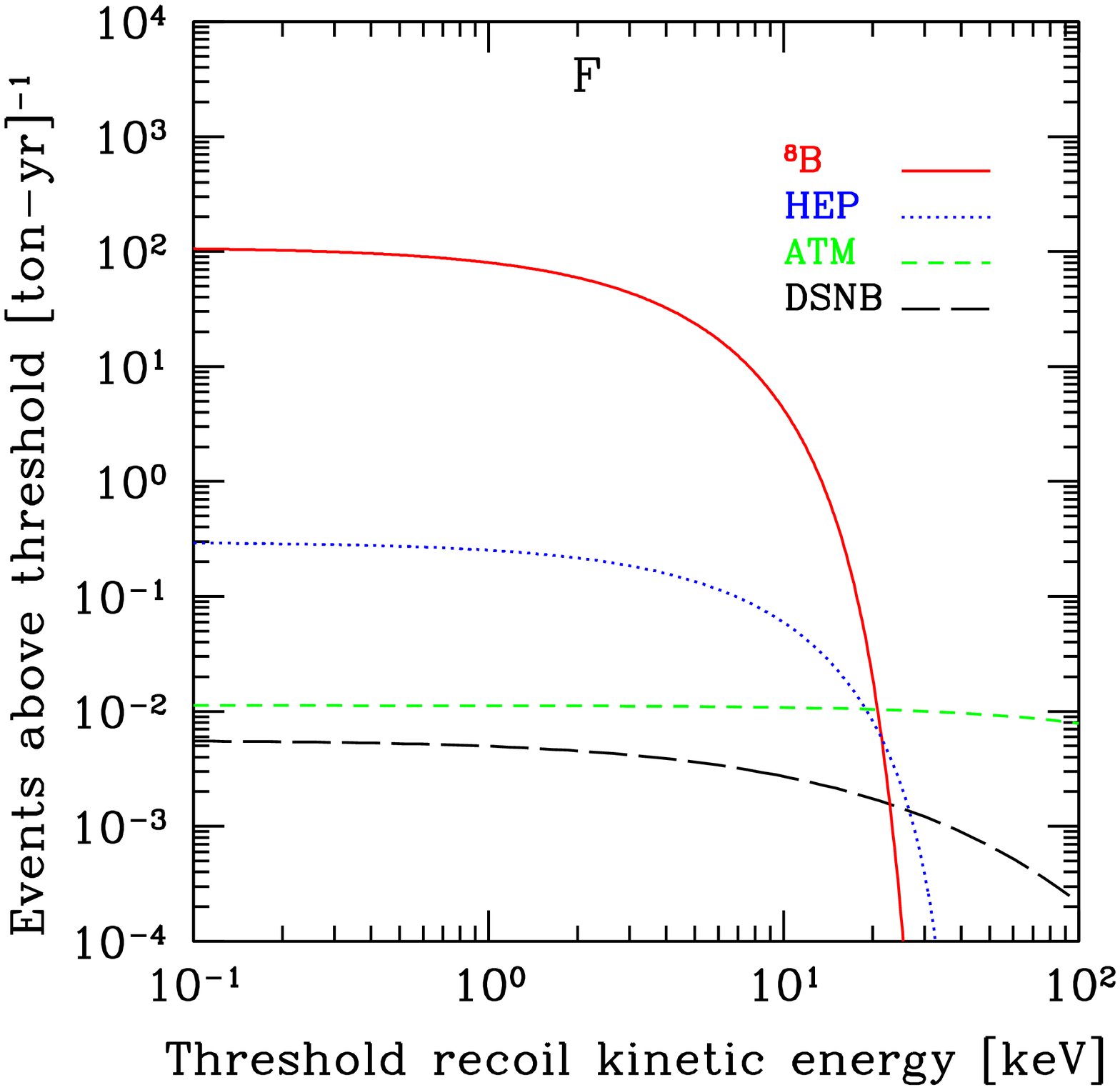} &
\includegraphics[height=7cm]{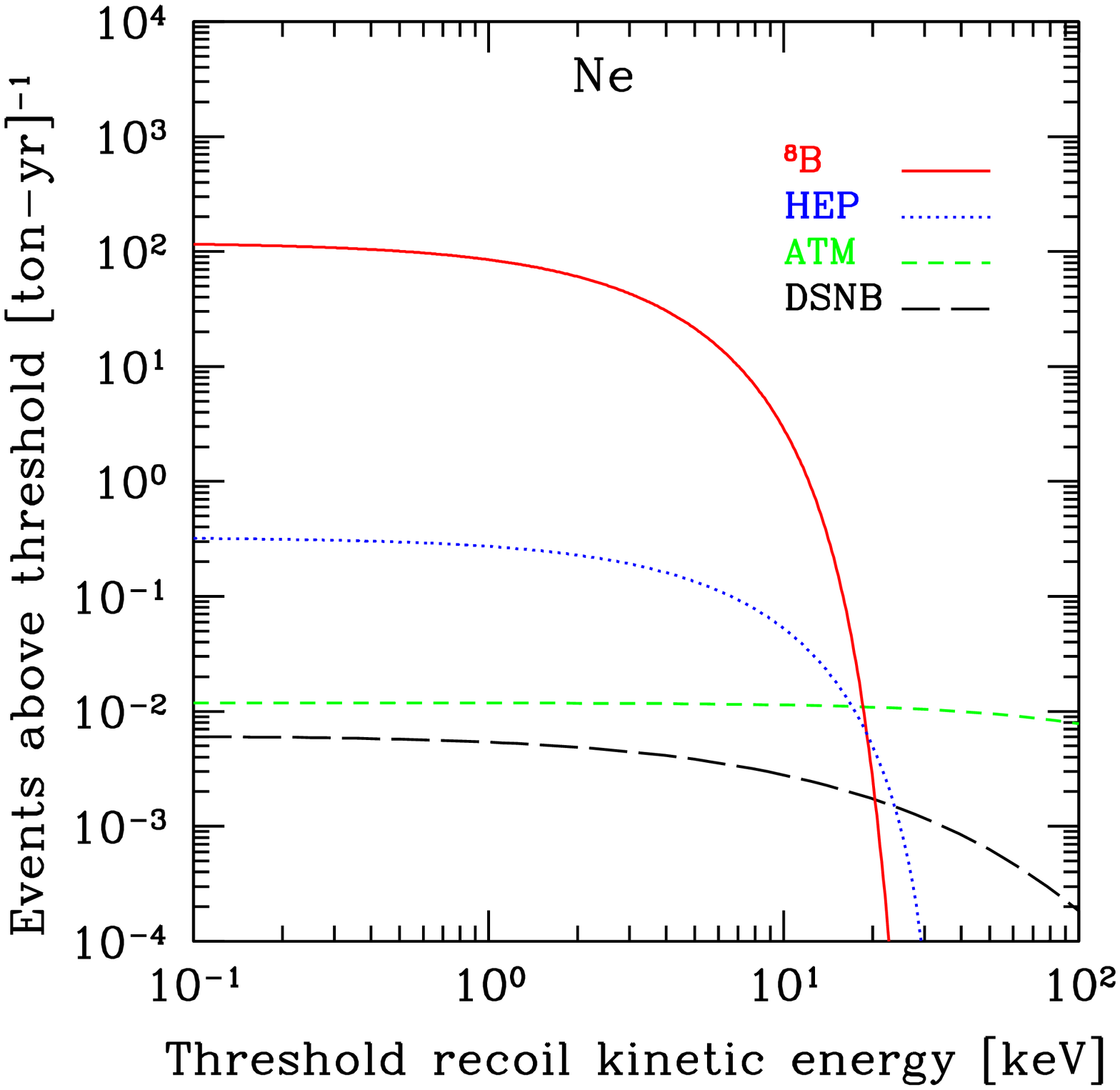} \\
\includegraphics[height=7cm]{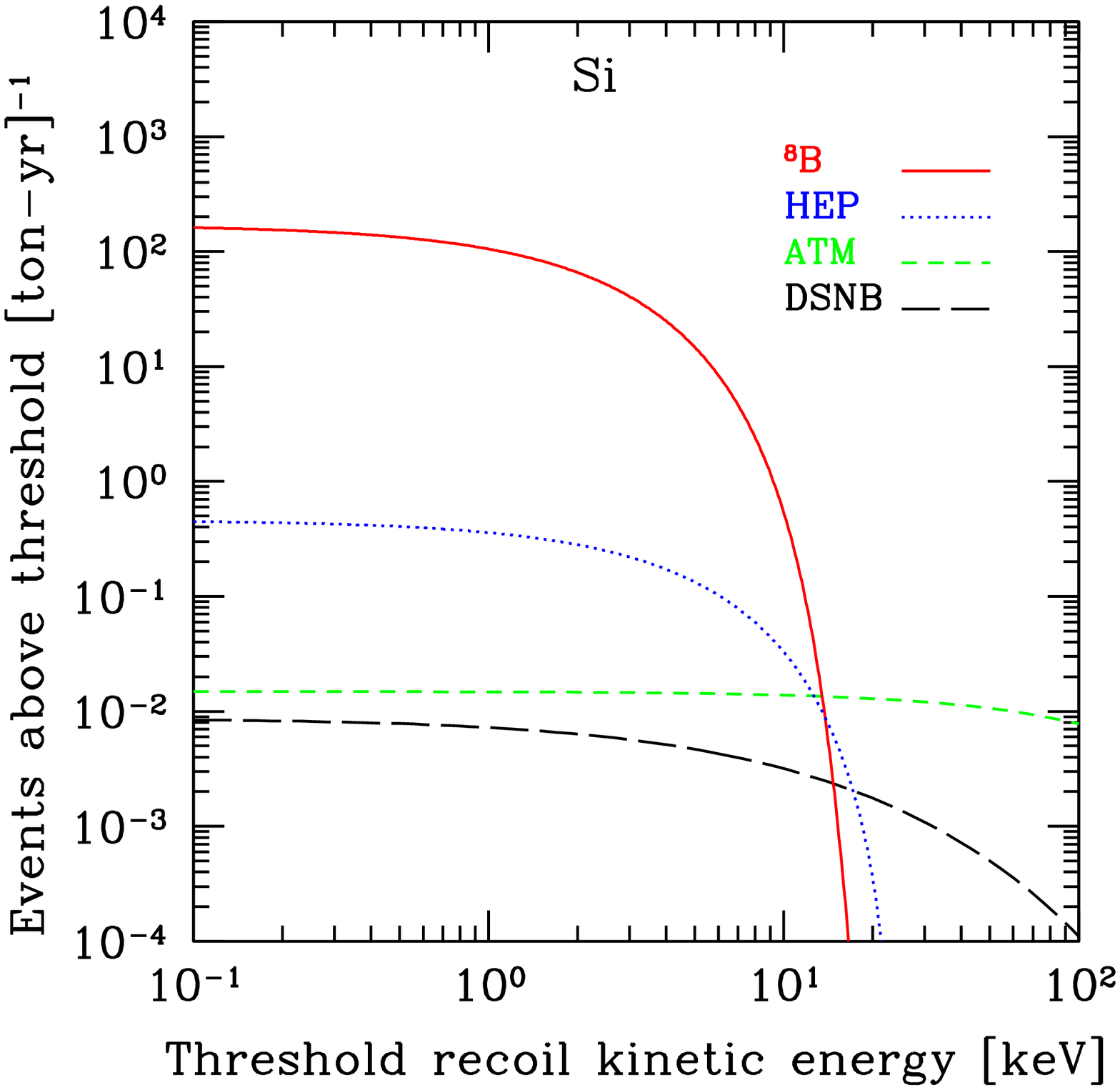} &
\includegraphics[height=7cm]{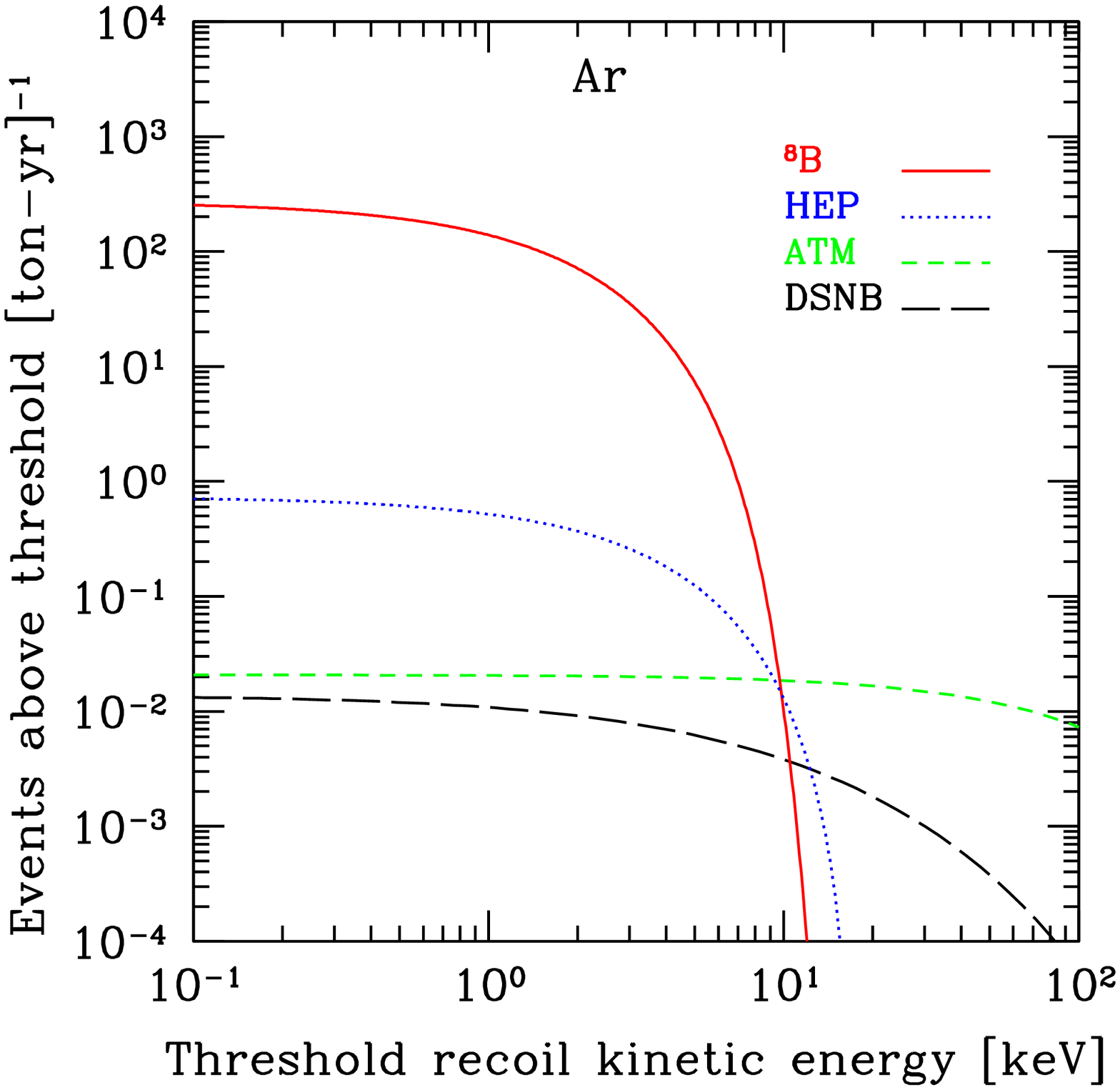} \\
\includegraphics[height=7cm]{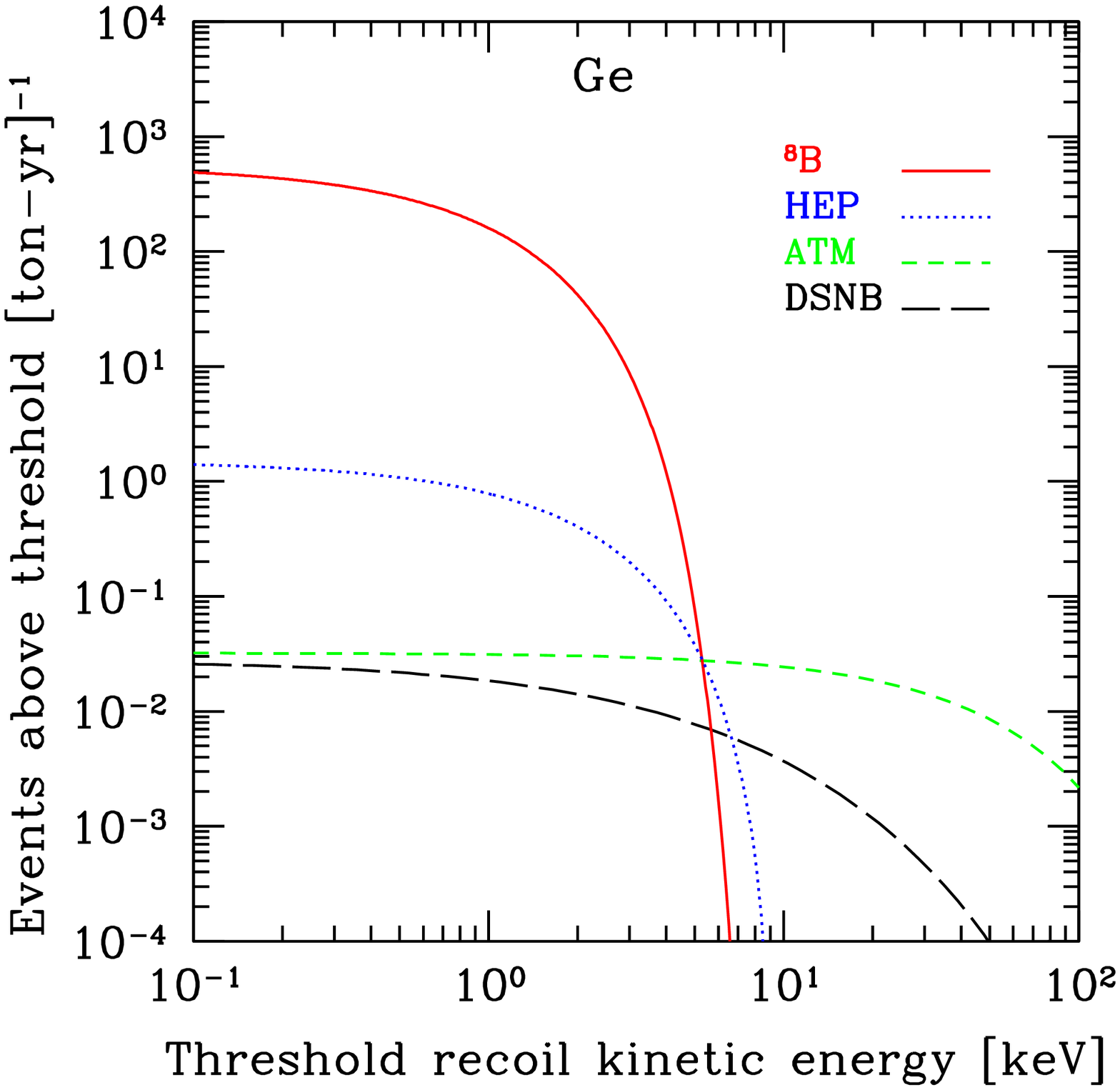} & 
\includegraphics[height=7cm]{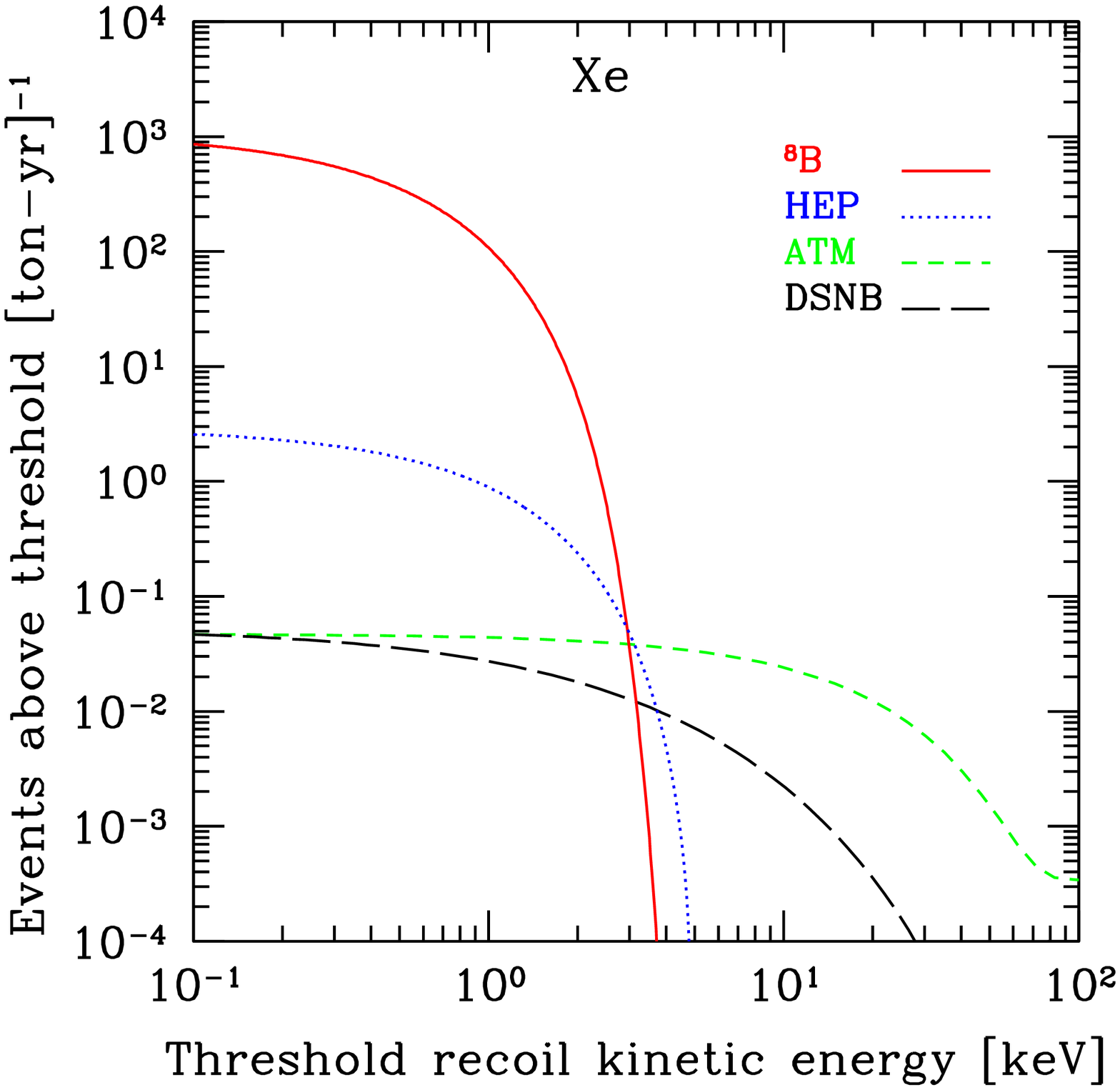} \\ 
\end{tabular}
\caption{Number of events above a threshold recoil kinetic energy 
for six target nuclei. For both the diffuse supernova and atmospheric event rates, the sum
of all contributing neutrino flavors are shown. 
\label{fig:threshold} 
}
\end{center}
\end{figure*}
\subsection{Event Rates} 

The differential event rate at a fixed recoil kinetic energy is 
\begin{equation}
\frac{dR(T)}{dT} = \int_{E_{\rm min}}^\infty \frac{dN}{dE_\nu} 
\frac{d\sigma(E_\nu,T)}{dT}dE_\nu, 
\label{eq:eventrate} 
\end{equation}
where $dN/dE_\nu$ is the neutrino flux spectrum and 
$E_{\rm min} = \sqrt{MT/2}$ is the minimum neutrino 
energy for a given recoil energy as dictated by the kinematics. 
Figure~\ref{fig:rates} shows the event rate recoil spectrum for 
six different nuclear targets. For all targets the
naturally-occurring abundances are assumed. For both the diffuse supernova and 
atmospheric event rates, the sum of all contributing neutrino flavors are shown. 
In particular for the {\it DSNB},  an 8 MeV spectrum from Figure~\ref{fig:fluxes} is
multiplied by four to account for the production spectrum of the four $\nu_x$ flavors. 
Due to their relatively hard spectra, the $\nu_x$ flavors are seen to dominate the event rate, particular 
at high recoil energies; there is only about a $\sim 10\%$ increase by
including the 3 and 5 MeV spectra at the lowest recoil energies. 
Each of these curves are the true, infinite resolution 
spectra, i.e. they do not account for the expected finite energy resolution 
of detectors. A detailed convolution with a resolution function will depend 
on the nuclear target and the particular experimental environment. 

Figure~\ref{fig:threshold} shows the number of events above
a given recoil energy for the same six nuclear targets. Most of
the $^8$B events are confined to low recoil energies, for
example for the case of Xe there are a total of $\sim 10^3$
events over all energies, but only $\sim 1$ event per ton-yr 
above a threshold of 3 keV. Future Xe detectors are expected
to have thresholds in the area of $\sim 5$ keV; as is seen dropping
the threshold below this energy will lead to a significantly increased
$^8$B signal. 

As an additional note, the analysis above just accounts for neutrino-nucleus
coherent scattering. In principle it would also be possible to detect
these same fluxes via neutrino-electron elastic scatterings~
\cite{Cabrera:1984rr}. For this channel the largest rate is to due the
solar {\it pp} reaction. For example, from {\it pp} scatterings on 
Xe a flat spectrum of 
electron recoil events is expected at $\sim 0.1$ events per ton-yr 
with energies up to $\sim 600$ keV. 

\section{Implications for WIMP-Nucleon Cross Section Constraints} 
In the absence of backgrounds the expected upper limit on the 
WIMP-cross section simply scales linearly with the
detector. For example a ten times
greater exposure will imply a ten times stronger
upper limit on the cross section. In the presence of 
backgrounds, however, the projected limits on the
cross section must be modified. 
Dodelson~\cite{Dodelson:2008yx} has provided a simple formalism for
estimating the upper limit on the WIMP-nucleon cross section, 
given a measured background rate and a fiducial detector volume. 
In this formalism, the probability of 
observing a total of $N$ events, given a WIMP-nucleon
cross section, $\sigma$, is 
\begin{eqnarray} 
{\cal L}(N|\sigma) 
\propto \int_0^\infty dN_b 
\exp\left[ \frac{-(N_b-\bar{N}_b)^2}{2\sigma_b^2}\right] 
\frac{e^{-\mu} \mu^N}{N!} . 
\label{eq:likelihood}
\end{eqnarray} 
Here the mean, $\mu$, is a sum of the WIMP-induced signal  
and neutrino-induced background events, and 
depends on the model parameters such as 
$\sigma$ and the expected background
spectrum. The mean number of background events is $\bar{N}_b$
and the error on the background is $\sigma_b$. 
In Eq.~\ref{eq:likelihood} the effect of binning
the data in energy is neglected, and as a result the estimates quoted
are likely to be conservative (binning the data would 
likely increase the sensitivity). 

From the likelihood in Eq.~\ref{eq:likelihood},~ Dodelson~\cite{Dodelson:2008yx}
defines the {\it Background
Penalty Factor} ({\it BPF}) as
\begin{equation} 
\sigma_{\rm future}^{\rm upper \, limit} = 
\frac{\sigma_{\rm current}^{\rm upper \, limit}}{\lambda} \times BPF, 
\label{eq:bpf}
\end{equation}
where $\lambda$ is the ratio  
of the future detector exposure to the current exposure. The {\it BPF} then essentially
measures, for a given background rate, the projected 
upper limit on the cross section for a future detector.  
It is calculated by integrating the normalized version of Eq.~\ref{eq:likelihood}
up to the desired confidence level, and the projected cross section 
upper limit 
is then obtained from Eq.~\ref{eq:bpf}. Here the projected upper 
limits are quoted at 95\% c.l.

The results presented here assume that the mean background
event rates $\bar{N}_b$
are obtained from the results in the figures above. The error
on the background is taken to be $\sigma_b = 0.2\bar{N}_b$, motivated by the uncertainty
on the solar neutrino fluxes~\cite{Bahcall:2004fg}. The 
results presented here are found 
to be independent of similar reasonable assumptions 
for the error on the background. As far as the actual measured 
backgrounds are concerned, 
recent results of XENON10~\cite{Angle:2008we} and 
CDMS~\cite{Ahmed:2008eu} give background
rates per unit volume of (7 events)/(0.44 kg-yr) and 
(0.6 events)/(0.33 kg-yr), respectively. These backgrounds
are not accounted for in this present analysis, 
amounting to the assumption that they 
can be reduced and understood in future experimental analysis. 
The calculations above 
show that, given the current exposures, the total background rate from all neutrino
sources is $\sim 10^{-4}$ events above CDMS and XENON10 thresholds,  
so it is unlikely that these
background events are neutrino-induced. 

\begin{figure*}[htbp] 
\begin{center}
\begin{tabular}{cc}
\includegraphics[height=7cm]{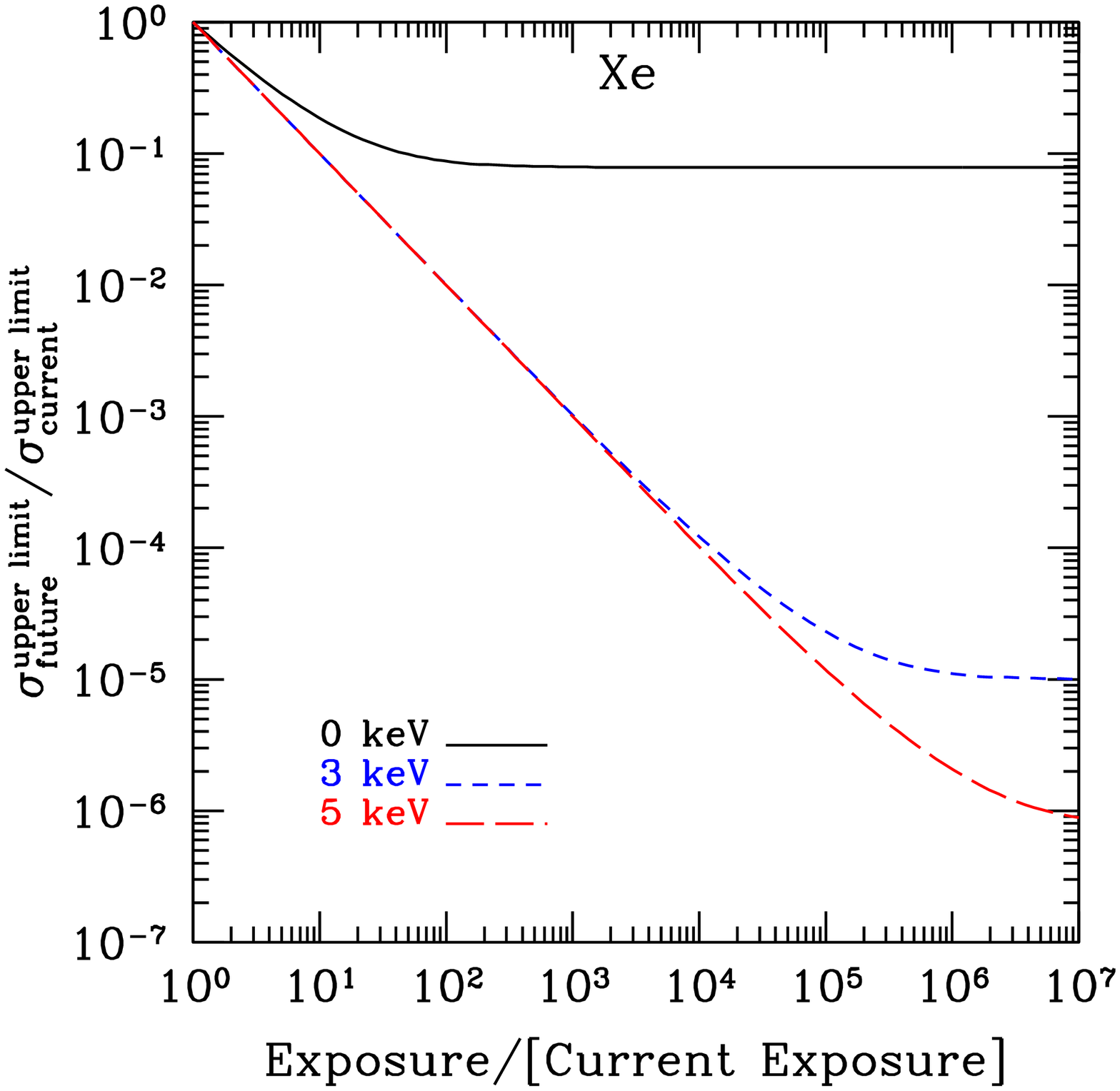} &
\includegraphics[height=7cm]{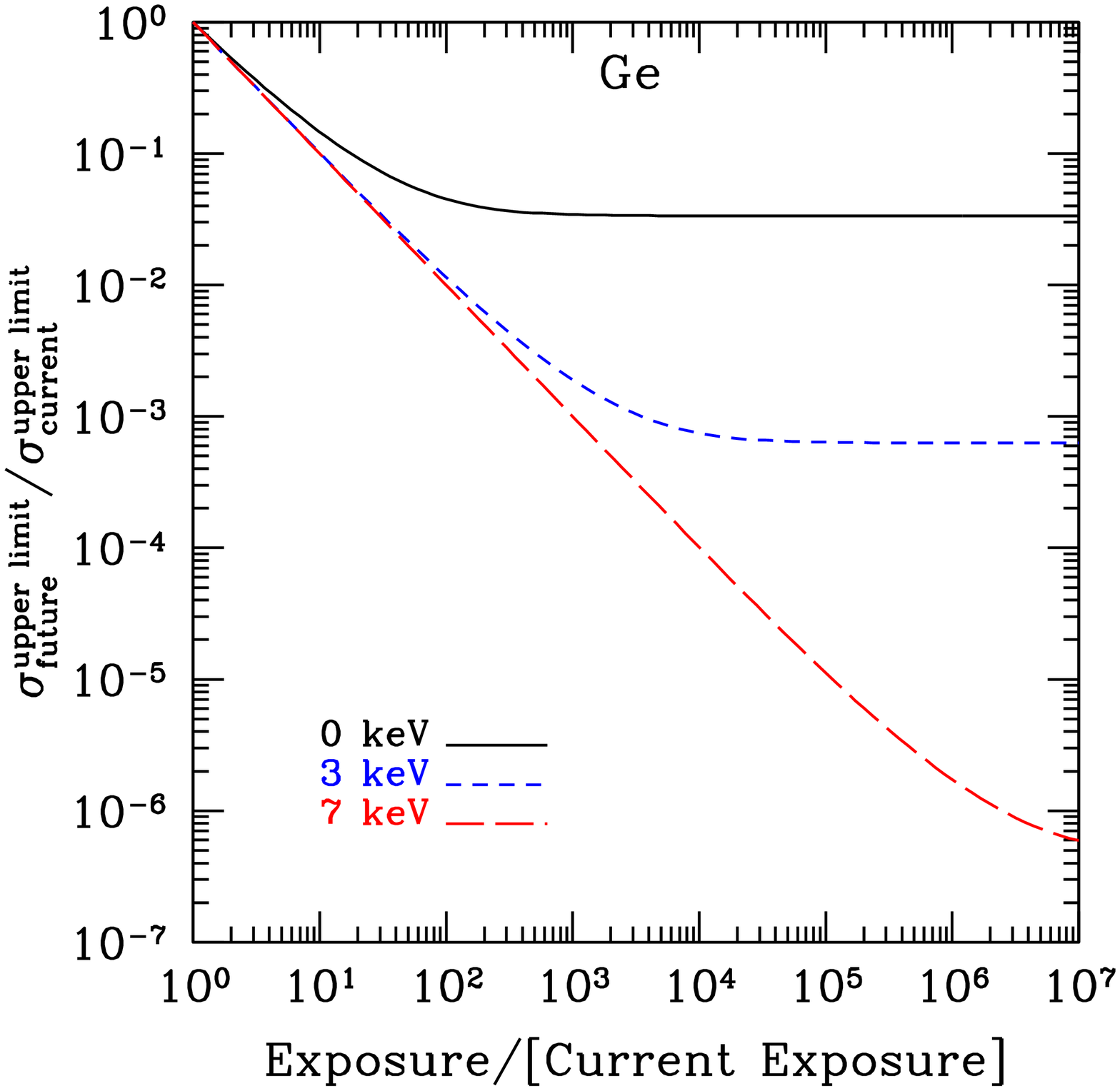} \\
\end{tabular}
\caption{The upper limit on the WIMP-nucleon cross section, 
relative to the current upper limit, as a function of the 
exposure, relative to the current exposure. Two target nuclei 
are shown, Xe and Ge. The labels in each case denote the
lower energy threshold assumed for the experiment. It is assumed
that there are no non-neutrino sources contributing to the background. 
\label{fig:bpf} 
}
\end{center}
\end{figure*}

Accounting for the neutrino backgrounds, 
Figure~\ref{fig:bpf} shows the upper limit on the cross section
attainable as a function of detector exposure. 
Two target nuclei are considered, Xe and Ge. To set the absolute scale, the current 
WIMP-nucleon cross section upper limits from these experiments 
are $\sim 4 \times 10^{-44}$ cm$^2$ (at $\sim 30$ GeV) 
and $\sim 5 \times 10^{-44}$ cm$^2$ (at $\sim 60$ GeV), 
respectively
~\cite{Angle:2008we,Ahmed:2008eu}. Each target shows the results for
three different energy thresholds. The salient point is that, as the
threshold is increased, the dominant $^8$B background is reduced
and the projected upper limit on the cross section will become much 
more stringent. In fact, cutting out the $^8$B signal entirely 
(which amounts to adopting the 5 keV threshold for Xe and the 7 keV threshold
for Ge), the projected upper limit will continue to decrease 
 to below $\sim 10^{-12}$ pb, the lower regime of the supersymmetric
 model parameter space~\cite{Roszkowski:2007fd}. 

It should be noted that, while effective at eliminating backgrounds,
setting the threshold at too high of an energy 
may have the adverse effect of cutting out the sought after signal events from
WIMP-induced recoils. 
For example, for a 100 GeV WIMP, $\sim 10\%$ of the recoil events occur in
energy bins below 3 keV. Similarly for a 50 GeV WIMP, $\sim 15\%$
of the events occur in energy bins below 3 keV. 
So setting the threshold to cut the entire $^8$B rate may
not be the optimal strategy, depending on the underlying WIMP parameters. 
In this sense, determining the WIMP signal robustly will require a
multi-component fit to both the WIMP and neutrino-induced recoil spectra. 

\section{Upper DSNB flux limit on non-electron neutrino flavors} 
\label{sec:upperlimit}
The preceding section has discussed the extent to which
the neutrino backgrounds can degrade the upper limits on the
WIMP-nucleon cross section. It is also of course interesting to explore 
the neutrino fluxes in the context of a signal themselves. 
The focus here is on the {\it DSNB} flux, and more specifically 
to provide an estimate for the expected flux limit that current detectors 
can set on the $\nu_x$ flux component. 
Similar to the preceding section, the projections focus on Xe and Ge targets, though they can Ê
easily be extended to other targets with similar results. 

From Figure~\ref{fig:threshold}, the expected
{\it DSNB} event rate is $\sim 0.05$ events per ton-yr. As
discussed above, this event rate is dominated by the four $\nu_x$
flavors, again not accounting for neutrino mixing. It follows that the detection of 
this flux requires an exposure of the order
100 ton-yr. To set the current scale, given that the current exposure for Xe is $0.44$ kg-yr, the
expected number of {\it DSNB} events is a paltry $\sim 2 \times 10^{-5}$. 

Nonetheless, from these predicted event rates 
an estimate of the flux upper limit that present detectors can 
set may be obtained by simply increasing the flux normalization so that ${\cal O}(1)$
event would be expected in detectors with the present exposures. The implied increase in the
flux normalization is $\sim 1/(2 \times 10^{-5}) = 5 \times 10^4$. 
Given that the current best estimates for the {\it DSNB} flux are $\sim 10$ cm$^{-2}$ s$^{-1}$
for non-electron flavors (again not accounting for neutrino mixing effects), 
it is projected that, with current exposures,  the $\nu_x$ upper flux limit will be reduced by 
more than an order of magnitude, so that the projected upper flux limit is 
\begin{equation}
\phi_{\nu_x}^{\rm limit} \, \ltsim \, 10^6 \, {\rm cm}^{-2} \, {\rm s}^{-1}. 
\label{eq:predicted}
\end{equation}
This limit is valid for all energies, so in this way it is better than an order of
magnitude greater than the Mt. Blanc flux limit, which is only valid
for neutrino energies $\gtsim \, 20$ MeV. 

The flux upper limit in Eq.~\ref{eq:predicted}
may be viewed as optimistic, because 
no contaminating backgrounds in the energy window of the {\it DSNB}  
signal have been accounted for. Accounting for the current backgrounds may somewhat degrade
this limit, though as in the preceding section it is assumed that these backgrounds
can be eliminated in future runs of the experiment. 
Going further and pushing the flux sensitivity even lower requires subtracting
the solar and atmospheric backgrounds. 
In fact, in order to determine how much better
the flux limit in Eq.~\ref{eq:predicted} will get as a function of detector exposure, 
it is possible to
again appeal to the {\it BPF} formalism from the previous section. 
The only difference is that $\phi_{\nu_x}^{\rm limit}$ replaces 
the cross section as the quantity to constrain. 

Under the assumptions above the projected upper limits on the 
{\it DSNB} flux are obtained simply by reading off the limits from
the curves in Fig.~\ref{fig:bpf}. Since the {\it DSNB} constitutes a small 
fraction of the total backgrounds in Fig.~\ref{fig:bpf},  it is found 
that calculating the flux limit in this manner provides an excellent approximation to
the case in which the curves in Fig.~\ref{fig:bpf} are calculated 
{\it without} the {\it DSNB} flux included. To set the normalization, 
the current flux limit is taken to be the projected value of $\phi_{\nu_x}^{\rm limit}$
from Eq.~\ref{eq:predicted}. For a 0 keV threshold
using an Xe target, from Fig.~\ref{fig:bpf} the upper flux limit is projected to
be reduced by about an order of 
magnitude further to $\ltsim \, 10^5$ cm$^{-2}$ s$^{-1}$ for a two order of 
magnitude increase in exposure. 
This scaling is again primarily because of the $^8$B solar neutrino background. 
Eliminating the $^8$B background with an energy threshold cut, Fig.~\ref{fig:bpf}
shows that the flux sensitivity will increase linearly with exposure
down to the projected {\it DSNB} flux window of $\sim 10$ cm$^{-2}$ s$^{-1}$
for exposures $\sim 10^5$ times the current exposures. 

Finally, it should be noted again that as in the case for the projected cross 
section limits, the {\it DSNB} flux estimates have 
not accounted for the detailed spectral shape of the signal or backgrounds. 
Only the energy threshold has been varied to eliminate the dominant $^8$B
background. Future detectors, with orders of magnitude more exposure than
at present should be able to improve on the upper limits provided here 
by using the difference between the spectral shapes of the signal and 
background. 

\section{Discussion and Conclusion} 
This paper has discussed the neutrino-induced 
recoil spectrum at dark matter detectors due to  
solar, atmospheric, and diffuse supernova neutrinos. 
It is shown that ton-scale detectors can expect $\sim 100$ $^8$B neutrino
events per ton year over all recoil energies. The dominant $^8$B signal
can be excluded using either spectral information or by 
an energy threshold cut. It should also be noted 
that the $^8$B signal could be cleanly identified by a detector with
directional sensitivity~\cite{Henderson:2008bn}. 

The effect of these neutrino-incuded
recoils on the extraction of an upper limit on the WIMP-nucleon
cross section has been examined. If the $^8$B signal can be excluded, and for the
particular case of Xe and Ge targets, the upper limit
on the cross section is projected to scale linearly with the detector
exposure. Non- $^8$B backgrounds only impact the WIMP-nucleon cross section
upper limit if the cross section is below $\sim 10^{-12}$ pb. 

\section*{Acknowledgments} 
I  thank Dan Akerib for discussions that motivated this paper.
I additionally thank John Beacom, Jodi Cooley-Sekula, Rick Gaitskell, 
Tom Shutt, Steven Yellin, and Hasan Yuksel for discussions and comments, as
well as Scott Dodelson for making his background penalty factor code
publicaly-available. 
Support for this work was provided by NASA through Hubble Fellowship grant 
HF-01225.01 awarded by the Space Telescope Science Institute, which is 
operated by the Association of Universities for Research in Astronomy, Inc., 
for NASA, under contract NAS 5-26555.

\vspace{2cm}

\bibliography{dm_bkgds}

\end{document}